\documentclass[reprint, superscriptaddress, prl]{revtex4-1}

\pdfoutput=1
\usepackage{mathtools, amsfonts, amsthm, latexsym, amssymb,dsfont}
\usepackage[T1]{fontenc}
\usepackage{microtype}

\usepackage{hyperref}
\hypersetup{colorlinks=true, citecolor=blue, urlcolor=blue}

\renewcommand{\d}{\partial}
\renewcommand{\O}{\mathcal{O}}
\renewcommand{\L}{\mathcal L}

\newcommand{\bra}[1]{\langle #1 \vert}
\newcommand{\ket}[1]{\vert #1 \rangle}

\newcommand{\matrixel}[3]{\langle #1 \vert #2 \vert #3 \rangle}
\newcommand{\vev}[1]{{\langle #1 \rangle}}
\newcommand{\del}{\nabla}

\usepackage{color}

\begin{document}
\title{Rotating superfluids and spinning charged operators in conformal field theory}
\author{Gabriel Cuomo}
\affiliation{Theoretical Particle Physics Laboratory, Institute of Physics, EPFL, Lausanne, Switzerland}
\author{Anton de la Fuente}
\affiliation{Theoretical Particle Physics Laboratory, Institute of Physics, EPFL, Lausanne, Switzerland}
\author{Alexander Monin}
\affiliation{Laboratory of Particle Physics and Cosmology, Institute of Physics, EPFL, Lausanne, Switzerland}
\author{David Pirtskhalava}
\affiliation{Theoretical Particle Physics Laboratory, Institute of Physics, EPFL, Lausanne, Switzerland}
\affiliation{Theoretical Physics Group, Blackett Laboratory, Imperial College, London SW7 2AZ, UK}
\author{Riccardo Rattazzi}
\affiliation{Theoretical Particle Physics Laboratory, Institute of Physics, EPFL, Lausanne, Switzerland}
\date{\today}

\begin{abstract}
We calculate the scaling dimensions of operators with large global charge and spin in 2+1 dimensional conformal field theories. By the state-operator correspondence, these operators correspond to superfluids with vortices and can be systematically studied using effective field theory. As the spin increases from zero to the unitarity bound, the superfluid state corresponding to the lowest dimension operator passes through three distinct regimes: (1) a single phonon, (2) two vortices, and (3) multiple vortices. We also calculate correlation functions with two such operators and the Noether current.
\end{abstract}
\maketitle

Perhaps a way to phrase the main difference between high energy physics and condensed matter physics is that high energy physics mostly occurs in the vacuum while condensed matter physics occurs at finite density. In a conformal field theory (CFT), this particular difference disappears: The state-operator correspondence maps finite density states to local operators and therefore maps finite density correlators to vacuum correlators. 

Recently, this idea was applied using superfluids \cite{hellerman1, alvarezGaume, riccardo, loukas, hellerman2, hellerman3, monteCarlo, matrixModels, hellerman4}. Superfluids are finite density states and correspond to operators with large global charge. Furthermore, superfluids are described by an effective field theory (EFT), allowing the computation of correlators in a systematic perturbative expansion \cite{son}. This EFT was used to study the CFT operator spectrum at large global charge~\cite{hellerman1, alvarezGaume, riccardo, loukas, hellerman2, hellerman3, monteCarlo, matrixModels, hellerman4}. 

In an independent line of research, much work was devoted to using the conformal bootstrap \cite{slavaCFT, dsdCFT} to study the CFT operator spectrum at large spin \cite{lightcone1,lightcone2,lightcone3,lightcone4,lightcone5,lightcone6,lightcone7,lightcone8,lightcone9,lightcone10,lightcone11}. This motivated us to ask the following question: Can large spin operators also be studied using EFT techniques? An obvious approach to this question is to start with the large charge operators studied by \cite{hellerman1, alvarezGaume, riccardo, loukas, hellerman2, hellerman3, monteCarlo, matrixModels, hellerman4} and then proceed by adding increasing amounts of spin to them. This translates to adding angular momentum to the corresponding superfluid. 

Experimentally, when angular momentum is added to a superfluid in the laboratory, vortices develop \cite{vortexBook}.
However, the superfluid EFT used by~\cite{hellerman1, alvarezGaume, riccardo, loukas, hellerman2, hellerman3, monteCarlo, matrixModels, hellerman4} does not incorporate vortices; all angular momentum is carried by phonons alone. This suggests that the EFT would incorrectly describe high angular momentum states. Conveniently, a superfluid EFT that incorporates vortices was recently constructed~\cite{nicolis}. We will use this EFT to study operators that have large spin as well as large charge.  

\paragraph*{Results.}
We found that we can study large charge operators with arbitrary spin, provided that the spin is parametrically below the unitarity bound. This range of spins is disjoint from the range in which the large spin bootstrap work \cite{lightcone1,lightcone2,lightcone3,lightcone4,lightcone5,lightcone6,lightcone7,lightcone8,lightcone9,lightcone10,lightcone11} is valid---that work only applies to operators parametrically close to the unitarity bound. Therefore, our results complement the bootstrap results.

We calculated the dimension $\Delta$ of the lowest dimension operator with charge $Q \gg 1$ and spin~$J$. As $J$ varies from~0 to $\Delta$, the corresponding superfluid state passes through three qualitatively distinct regimes. We will simply state the results now and derive them later. The results are displayed at leading order in both large~$Q$ and large~$J$. At this order, there is only a single free parameter~$\alpha$ for the entire range of $J$. 

For $0 \leq J \lesssim \sqrt Q$, the lowest energy state has no vortices and consists of a single phonon of angular momentum $J$. The corresponding operator dimension $\Delta$ is \cite{hellerman1, riccardo}
\begin{equation} \label{results1}
	\Delta= \alpha Q^{3/2} + \frac{J}{\sqrt 2} 
\end{equation}

For $\sqrt Q \lesssim J \leq Q$, the lowest energy state consists of a vortex-antivortex pair whose separation increases with~$J$. The corresponding operator dimension $\Delta$ is
\begin{equation} \label{results2}
	\Delta =  \alpha Q^{3/2} +  \frac{\sqrt Q}{3\alpha} \ln \frac{J}{\sqrt Q}.
\end{equation}

For $Q < J \lesssim Q^{3/2}$, the lowest energy state consists of multiple vortex-antivortex pairs distributed so that the superfluid has the same velocity profile as that of a rotating rigid body \cite{feynman}. The corresponding operator dimension~$\Delta$ is
\begin{equation} \label{results3}
	\Delta = \alpha Q^{3/2} +  \frac{1}{2 \alpha}  \frac{J^2}{Q^{3/2}}.
\end{equation}

As $J \to Q^{3/2} \sim \Delta$, the EFT breaks down and, as mentioned, we are unable to reach the spin of the operators studied in \cite{lightcone1,lightcone2,lightcone3,lightcone4,lightcone5,lightcone6,lightcone7,lightcone8,lightcone9,lightcone10,lightcone11} by bootstrap methods. 

Our results apply to any CFT that satisfies three conditions: first, its large charge sector can be described as a superfluid; second, this superfluid admits vortices; third, the only low energy degrees of freedom are the Goldstone modes of the superfluid.
These are the simplest and most natural conditions we can imagine. Because of this, we believe---but cannot prove---that our results apply to a wide range of CFTs with a $U(1)$ global symmetry. For example, we expect that they apply to the critical $O(2)$ model \cite{wilsonFisher} and can be tested in principle. 

\paragraph*{General Strategy.} 
We will now explain in more detail how the above results were derived. 
Our general strategy consists of combining two powerful tools: effective field theory~(EFT) and the state-operator correspondence. We begin by considering a $d$+1 dimensional CFT on a cylinder $\mathbb R \times S^d$. Next, we assume that a given EFT on this cylinder is a valid description of our CFT. Finally, we apply the state-operator correspondence directly to the states of this EFT. 
Throughout the paper, unless otherwise stated, we shall work at leading order in the derivative and field expansion within EFT \cite{hellerman1,riccardo}.

Note that this strategy differs from that in~\cite{riccardo}, which takes the ``top-down''  approach of projecting onto a desired state using the Euclidean path integral. Instead, we are taking the ``bottom-up'' approach of simply assuming an EFT and then quantizing its Hamiltonian while always remaining in Lorentzian spacetime.

As a reminder, in a CFT, there is a one-to-one correspondence between the eigenstates of the Hamiltonian~$H$ on~$S^{d}$ and the set of scaling operators at any given point. This is called the state-operator correspondence (see \cite{slavaCFT, dsdCFT} for reviews). The energy $E$ of a state is related to the scaling dimension $\Delta$ of the corresponding operator by $E = \Delta/R$, where $R$ is the radius of the sphere~$S^d$. All other conserved quantum numbers of the state (such as global charge and spin) are identical to those of the corresponding operator.

\paragraph*{Dual gauge field.}
We now specialize the construction of \cite{nicolis} to the cylinder $S^2 \times \mathbb R$. It uses a dynamical gauge field $a_\mu$ instead of the more familiar Goldstone field~$\pi$.
We begin with the conformal superfluid effective Lagrangian \cite{son, hellerman1, riccardo} written in terms of $\pi$:
\begin{equation} \label{superfluid}
	\L= c (\d \chi)^3,
\end{equation}
where $\chi \equiv \mu t + \pi$, $c$ is an unknown constant, and $\mu$ can be interpreted as the chemical potential. We use the notation, $w \equiv (g^{\mu\nu}w_\mu w_\nu)^{1/2}$, where $g_{\mu\nu}$ is the spacetime metric and~$w_\mu$ is an arbitrary spacetime vector. We then dualize $\chi$ by formally treating $v_\mu \equiv \d_\mu \chi$ as an independent variable and introducing a Lagrange multiplier~$a_\mu$ to set the curl of $v_\mu$ to zero:
\begin{equation}
	\L = c\, v^3 - \frac{1}{2\pi} a_\mu \frac{\epsilon^{\mu\nu\lambda}}{\sqrt g} \d_\nu v_\lambda .
	\label{av_lagrangian}
\end{equation}
Note that in our notation, it is the combination $\epsilon^{\mu\nu\lambda}/\sqrt{g}$ that gives the antisymmetric Levi-Civita \emph{tensor}.
Integrating out $v_\mu$ gives
\begin{equation} \label{gauge}
	\L = -\kappa f^{3/2},
\end{equation}
where $f \equiv \sqrt{f_{\mu\nu}f^{\mu\nu}}$ and $f_{\mu\nu} \equiv \d_\mu a_\nu - \d_\nu a_\mu$. We dropped a boundary term that came from integrating \eqref{av_lagrangian} by parts because it is metric independent and thus does not affect the energy momentum tensor. The coefficient $\kappa$ in \eqref{gauge} is related to the coefficient $c$ in \eqref{superfluid} as $\kappa = \frac{1}{2^{5/4}(3\pi)^{3/2}} \frac{1}{\sqrt c}$.

The relation between $\chi$ and $a_\mu$ is given by the expression for the $U(1)$ current $j^\mu$:
\begin{equation} \label{current}
	j^\mu = 3 c (\d \chi) \d^\mu \chi = \frac{1}{4\pi} \frac{\epsilon^{\mu\nu\lambda}}{\sqrt g} f_{\nu\lambda} . 
\end{equation}
In the vacuum, the charge density is $\vev{j^0}=\frac{Q}{4\pi R^2}$, where~$Q$ is the net charge of the superfluid state and $R$ is the radius of the sphere. This translates to a homogeneous magnetic field
\begin{equation}\label{Bdef}
	\vev{f_{\theta\phi}} = B \sin \theta \equiv \frac{Q}{2R^2} \sin \theta
\end{equation}
and results in a net magnetic flux of $2\pi Q$ through the sphere. Parametrically, the cutoff $\Lambda$ of our EFT is
\begin{equation} \label{cutoff}
	\Lambda \sim \sqrt B \sim \frac{\sqrt Q}{R}.
\end{equation}

\paragraph*{Particle-vortex duality.}
Vortices in the superfluid description correspond to heavy charged particles in the gauge theory description (see \cite{zee} for a pedagogical introduction). They are treated in a first-quantized form as 0+1 dimensional worldlines embedded in the 2+1 dimensional spacetime. We will use the terms ``vortex'' and ``charged particle'' interchangeably.

To write the effective action for a  conformal superfluid with vortices  \cite{nicolis}, we parametrize the spacetime trajectory of the $p$th vortex  by  $X_p^\mu(\tau)$, where $\tau$ is an auxilliary time parameter. We further impose Weyl and $\tau$-reparametrization invariance, with the former reducing to conformal invariance in the relevant case of a static metric. The action can be organized as a derivative expansion with the lowest order terms given by 
\begin{multline}\label{action}
	S= -\kappa \int d^3 x \sqrt g \, f^{3/2} - \sum_p q_p \int a_\mu dX_p^\mu \\
	 - \sum_p  \int d\tau \, \gamma_p \sqrt f\sqrt{ g_{\mu\nu} \dot X_p^\mu \dot X_p^\nu}+\cdots  \,\,.
\end{multline}
The first term is the kinetic term \eqref{gauge} for the gauge field. The second term is the leading coupling between a particle of charge $q_p$   and the gauge field \cite{landauCTF}. The last term can be viewed as the action for a relativistic point particle~\cite{landauCTF} with mass~$\gamma_p \sqrt f$. 
The dots in \eqref{action} represent terms with at least two derivatives on either $A_\mu$ or $X_p^\mu$, and the coefficient~$\gamma_p$ should in general be promoted \cite{nicolis} to a function of $j_\mu \dot X^\mu / (j\dot X)$, where~$j_\mu$ is the $U(1)$ current \eqref{current}.
However, as we will now explain, the leading order description of our system is fully determined by the first line of \eqref{action}. We note that we are restricting ourselves to parity-invariant CFTs. Otherwise, we could have added a one-derivative Chern-Simons term, $A\wedge dA$, to the action~\eqref{action}.
\paragraph*{Lowest Landau Level.}
Each $X_p^\mu$ describes the motion of a 2D particle in a magnetic field and consists of two pairs of canonically conjugate variables. These can be further decomposed into one pair that describes the motion of the guiding center and another pair that describes cyclotron motion. Without interparticle interactions, excitations of the first pair are gapless while excitations of the second pair have a gap $\omega = B/m$, where $m$ is the particle mass. The gapped excitations are the Landau levels~\cite{landauQM}, and the gaplessness of the guiding center variables is  the usual  degeneracy of Landau levels.

In our system \eqref{action}, $\omega \sim \sqrt B$, which coincides with the EFT cutoff~\eqref{cutoff}. Thus, within the domain of validity of our EFT, the dynamics of $X^\mu$ reduces to that of just the guiding center---the Landau levels are effectively integrated out \cite{integrateLandau1,integrateLandau2,integrateLandau3}. 
One well-known fact \cite{jackiw1,jackiw2} is that the guiding center can be described by dropping the mass term (the second line) from the Lagrangian \eqref{action}.
Physically, this is because in the massless limit, the Landau level gap $\omega \to \infty$. Formally, this is because the first line in~\eqref{action} is \emph{linear} in the particle velocity. This constrains the two physical coordinates to be canonically conjugate to each other, halving the dimension of phase space. 

In view of the above, the terms in the second line of~\eqref{action} are genuine higher derivative corrections. On the one hand,
they bring in new states with energy~$\sim\sqrt B$. On the other, at sufficiently low energy, they can be consistently treated as small perturbations of the leading single derivative term. This is fully analogous to supplementing the 1D Lagrangian $\dot q^2 $ with the 4-derivative term $\ddot q^2/\Lambda^2$: states arise with energy $\sim \Lambda$, but at low-energy, the 4-derivative term can be treated as a perturbation using standard EFT methods. We leave a systematic study of higher order corrections for future work, though we will briefly mention some effects below.  

In what follows, we will therefore derive leading order results by
simply dropping the second line in~\eqref{action} and assume
\begin{equation}\label{actionNoLandau}
	S= -\kappa \int d^3 x \sqrt g \, f^{3/2} - \sum_p q_p \int a_\mu dX_p^\mu.
\end{equation}
The electrostatic potential coupling will generate nontrivial dynamics and a gap for the guiding centers. 
As long as particle separations are  larger than the cut-off length $1/\sqrt B$, this is within the regime of validity of the EFT.

\paragraph*{Classical Analysis.}
We will now compute the classical energy and angular momentum of a state with a given configuration of vortices. By the state-operator correspondence, this gives us the dimension and spin of the corresponding operator. We will work to leading order in large $Q$ and in large vortex separations.
At this order, the equations of motion from~\eqref{actionNoLandau} are
\begin{gather}
	\frac{1}{e^2} \del_\mu f^{\mu\nu} = \mathcal J^\nu \label{maxwell} \\
	E^i = (\dot X_p)_j  f^{ji}  \label{lorentz} ,
\end{gather}
where $E^i \equiv f^{i0}$ is the electric field and $\mathcal J^\nu$ is the current due to the point charges. The coupling $e^2$ is defined as
\begin{equation} \label{constants}
	\frac{1}{e^2} \equiv \frac{3\kappa}{2^{1/4}} \frac{1}{\sqrt B} .
\end{equation}
Eqs.~\eqref{maxwell} are Maxwell's equations and \eqref{lorentz} imposes that the particles move on trajectories with vanishing Lorentz force. In other words, as expected, the particles exhibit pure drift velocity motion. This is consistent  with cyclotron degrees of freedom being integrated out. Thus, the particle velocities are~$| \dot{  \vec X}_p | \sim | \vec E | /B  \sim 1/ \sqrt Q$
and can be neglected. Our problem has thus reduced to the 2D electrostatics of point charges on a sphere in a constant magnetic field.

The stress energy tensor $T_{\mu\nu} = \frac{2}{\sqrt g} \frac{\delta S}{\delta g^{\mu\nu}}$ is
\begin{equation}\label{stressTensor}
	T_{\mu\nu} = \frac{\kappa}{\sqrt f} \left(-3 f_{\mu\alpha} f_\nu{}^\alpha + g_{\mu\nu} f^2 \right).
\end{equation}
Using this, we calculate the dimension $\Delta$ of the corresponding operator:
\begin{equation} 
	\Delta =\frac{Q^{3/2}}{\sqrt{27\pi c}}  +\frac{R^3}{2e^2}\int d\theta d\phi \sin\theta  \vec E^2.  \label{fieldEnergy}
\end{equation}
Physically, the first term is the energy stored in the background magnetic field while the second term is the energy stored in the electric field created by the particles. 

In Coulomb gauge, $E_i = \d_i a_0$ where $a_0$ is the electric potential due to a collection of point charges on a 2-sphere:
\begin{equation} \label{potential}
	a_0(\vec r) = - \frac{e^2}{4\pi}  \sum_p q_p \ln (\vec r - \vec R_p)^2.
\end{equation}
We used embedding coordinates of the 2-sphere in $\mathbb R^3$, where $\vec r = (\sin \theta \cos \phi, \sin \theta \sin \phi, \cos \theta)$ is a unit 3-vector and analogously for~$\vec R_p$. The total electric field energy is a sum of pairwise contributions for each charge:
\begin{equation} \label{divergence}
	\frac{Re^2}{8\pi}\biggl[ -\sum_{p\neq r} q_p q_r \ln(\vec R_p - \vec R_r)^2 - \sum_p q_p^2 \ln 0^2 \biggr].
\end{equation}
The last term is the familiar divergent self-energy of a point charge. It will be cut off at angular lengths $\sim1/\sqrt Q $ \eqref{cutoff}. More formally, the divergence will be cancelled by the bare mass term we dropped in \eqref{actionNoLandau}, and the logarithm will come from renormalization group running \cite{nicolis}. 

Thus, the dimension $\Delta$ \eqref{fieldEnergy} is
\begin{equation} \label{dimension}
	\Delta =  \alpha Q^{3/2} - \frac{\sqrt Q}{12\alpha} \sum_{p \neq r} q_p q_r \ln \frac{(\vec R_p - \vec R_r)^2}{Q},
\end{equation}
where $\alpha \equiv 1/\sqrt{27\pi c}$, and we used $\sum_p q_p =0 $ to combine the logarithms in \eqref{divergence}. 

The angular momentum $\vec J$  can also be calculated from the stress tensor \eqref{stressTensor} and is
\begin{equation} \label{J}
	\vec J = -\sum_p q_p \frac{Q}{2} \vec R_p.
\end{equation}

\paragraph*{Derivation of results.}
The results stated at the beginning of this paper can now be derived. First, note that the self-energy of a particle of charge $q$ is proportional to $q^2$---this is the second term in \eqref{divergence}. Because of this, particles with $|q| > 1$ are energetically unfavored.

\noindent \textbullet \, Eq.~\eqref{results1} is derived using the phonon dispersion relation,~$\omega = [\tfrac12 \ell(\ell+1)]^{1/2}$ \cite{hellerman1,riccardo}, where $\ell$ is angular momentum and $\omega$ is energy. Since $\omega/\ell$ decreases with $\ell$,  energy is lowest at fixed $J$ with a single phonon of~$\ell = J$. 

\noindent \textbullet \, Eq.~\eqref{results2} is derived by evaluating $\Delta$ \eqref{dimension} and $J$ \eqref{J} on a configuration with a single vortex-antivortex pair. 

\noindent \textbullet \, Eq.~\eqref{results3} is derived by approximating the vortex distribution as a continuous distribution and then minimizing~$\Delta$~\eqref{dimension} for fixed $J$ \eqref{J} using variational techniques. This gives a vortex distribution $\rho$ of
\begin{equation} \label{vortexDensity}
	\rho = \frac{3}{2\pi R^2} \frac{J}{Q} \cos \theta
\end{equation} 
and results in the superfluid having the same velocity profile as that of a rigid body~\cite{feynman}.

\noindent \textbullet \, As $J \to Q^{3/2}$, the electric field $|\vec E|$ approaches the magnetic field $B$ and the drift velocities become relativistic. This causes the EFT to break down because the higher order terms neglected in \eqref{actionNoLandau} become unsupressed. The guiding centers becomes as energetic as the cyclotron degrees of freedom and anything else at the EFT cutoff~\eqref{cutoff}.

\paragraph*{Quantization.}
Since the vortex positions are continuous, some questions may occur: How many distinct states are there? How does the quantization of angular momentum arise? These questions are answered when we quantize our system of charged particles in a magnetic field. Solving for $a_0$ using \eqref{potential} and ignoring fluctuations of $a_i$, our effective Lagrangian \eqref{actionNoLandau} becomes
\begin{equation} \label{particleL}
	L = \sum_p q_p \vec A \cdot \dot{\vec R}_p + \frac{e^2}{8\pi} \sum_{p,r} q_p q_r \ln(\vec R_p - \vec R_r)^2,
\end{equation}
where $\vec A$ is the potential for a magnetic monopole~\cite{monopoleHarmonics, coleman}. We use the gauge in which $A_\phi = \tfrac12 Q(1-\cos \theta)$ and $A_\theta = 0$. This system is known as the ``fuzzy sphere''~\cite{fuzzySphere1,fuzzySphere2}. 

Due to the somewhat complicated form of $\vec A$, it is useful to switch to spinor coordinates \cite{haldane, greiter}: 
\begin{equation}
	\psi \equiv \begin{pmatrix} \cos \frac{\theta}{2} \\ \sin \frac{\theta}{2} e^{i\phi} \end{pmatrix},
\end{equation}
where we suppressed the vortex index. In these coordinates, $\vec R = \psi^\dag \vec \sigma \psi$ and $\vec A \cdot \dot{\vec R} = -i Q \psi^\dag \frac{d}{dt} \psi$.
This identifies the canonical momentum corresponding to $\psi_p$ as~$-iQq_p \psi_p^\dag$. The canonical commutation relations imply that the angular momentum \eqref{J} commutes with the Hamiltonian and satisfies $[{\mathcal J}_i, {\mathcal J}_j]~=~i \epsilon_{ijk} {\mathcal J}_k$~\cite{fuzzySphere1,fuzzySphere2}. We use curly $\vec{\mathcal J}$ to denote the angular momentum \emph{operator}.

For illustration, consider the case of two vortices of unit charge. The Hamiltonian corresponding to~\eqref{particleL} is then
\begin{equation} \label{Hvortex}
	H = \text{const} + \frac{e^2}{4\pi} \ln \vec {\mathcal J}^{\,2},
\end{equation}
where ``const'' involves terms that are independent of the vortex coordinates, and we used \eqref{J} to express $H$ in terms of $\vec {\mathcal J}$.  The spectrum is thus entirely determined by the spectrum of $\vec {\mathcal J}^{\,2}$. As is well-known, $\vec {\mathcal J}^{\,2} = J(J+1)$, where $J$ is an integer and for each value of $J$, there are $2J+1$ degenerate states. 

Restoring the constants in \eqref{Hvortex}, the dimension of the corresponding operator is
\begin{equation} \label{quantum}
	\Delta =  \alpha Q^{3/2} +  \frac{\sqrt{Q}}{6\alpha} \ln \frac{J(J+1)}{Q}.
\end{equation}
We can trust this equation for $\sqrt Q \lesssim J \leq Q$.
The lower value is determined by requiring that the vortices be separated by distances larger than the cutoff~$\sim 1/ \sqrt B$. The upper value occurs when the two particles are at opposite poles.

\paragraph*{Correlators.}
The EFT we presented can also be used to compute correlation functions \cite{riccardo}.
Let us consider correlators involving the $U(1)$ current $j_\mu$. From \eqref{current} and Gauss's law, we see that the line integral $\oint j_\mu dx^\mu$ about a closed curve~$\mathcal C$ at fixed time is simply $\frac{1}{2\pi}$ times the total charge~$q_\text{enc}$ enclosed by $\mathcal C$:
\begin{equation} \label{lineInt}
	\bra{\text{vortex}} \oint_{\mathcal C} j_\mu dx^\mu \ket{\text{vortex}} = \frac{e^2 q_\text{enc}}{2\pi},
\end{equation}
where $\ket{\text{vortex}}$ is a generic vortex state. By the state-operator correspondence, this amounts to a nontrivial prediction about three-point functions. We will now consider two simple examples.

As a first example, we consider a vortex-antivortex pair located at the north and south poles. Then~\eqref{lineInt} becomes
\begin{equation} \label{eft3pt}
	\matrixel{\text{vortex}}  {j_\phi(\theta,\phi)} {\text{vortex}} =  \frac{e^2}{2\pi R},
\end{equation}
where now $\ket{\text{vortex}}$ is a state with $J = J_z = Q$ and~$j_\phi$ is the azimuthal component of $j_\mu$.
In general, the expectation value of a spin-1 operator~$j_\phi$ in a state $\ket{J,J_z}$ with $J = J_z = Q$ is \cite{spinning} 
\begin{equation} \label{general3pt}
	\matrixel{Q,Q}{j_\phi(\theta,\phi)}{Q,Q} = R^2 \sum_{m=0}^Q a_m \cos^{2m}\theta,
\end{equation}
where $a_m$ are arbitrary (theory-dependent) constants subject to the constraint $\sum_m a_m=0$. By equating \eqref{eft3pt} to~\eqref{general3pt}, we obtain the following predictions for $a_m$ at leading order:
\begin{equation}
	a_m = 
	\begin{cases}
		\frac{\sqrt Q}{3\alpha}, &\text{if $m=0$;}\\
		0, &\text{if $1\leq m \ll\sqrt Q$.}
	\end{cases}
\end{equation}
We can only make predictions for $m\ll \sqrt{Q}$ because the EFT breaks at distances of order of the inverse cutoff~\eqref{cutoff} from the vortices. The constraint $\sum_ma_m=0$ is thus irrelevant for our discussion.

As a second example, we consider the states described by \eqref{results3}.  Using \eqref{vortexDensity}, we find
\begin{equation}
	\matrixel{\text{vortex}}  {j_\phi(\theta,\phi)} {\text{vortex}} =  \frac{3e^2}{8\pi^2R} \frac{J}{Q} \sin^2\theta.
\end{equation}
This expression is valid as long the continuous approximation for the density \eqref{vortexDensity}
can be used. In other words, it is valid on distance scales larger than the vortex separation 
$\sim1/\sqrt{\rho}\sim \sqrt{Q/J}$. Rewriting (\ref{general3pt}) in the Fourier basis:
\begin{equation}
	\matrixel{Q,Q}{j_\phi(\theta,\phi)}{Q,Q} = R^2 \sum_{m=0}^J b_m \cos 2m \theta,
\end{equation}
we obtain the following predictions for $b_m$ at leading order:
\begin{equation}
	b_m = 
	\begin{cases}
		\frac{(-1)^m}{8\pi \alpha} \frac{J}{\sqrt Q}, &\text{if $m=0,1$;}\\
		0, &\text{if $2\leq m \ll\sqrt{J/Q}$.}
	\end{cases}
\end{equation}

\paragraph*{Large $N$.} So far, we assumed that the numerical coefficients in our effective Lagrangian are $O(1)$, corresponding to an underlying strongly coupled CFT. 
The case of a weakly coupled or large $N$ theory is quickly illustrated. However, the conclusions depend on whether the weak coupling appears in the $\chi$ description~\eqref{superfluid} or in the $a_\mu$ description~\eqref{gauge}. We will refer to the $\chi$ description as ``electric'' and to the $a_\mu$ one as ``magnetic'', with couplings $g_e^2 \equiv 1/N_e$ and  $g_m^2 \equiv 1/N_m$, respectively.

Consider first a weakly coupled magnetic theory (for example, the large $N_m$ setup discussed in \cite{largeN1,largeN2,largeN3,largeN4,largeN5,largeN6,largeN7,largeN8}). 
Since $1/\alpha\sim  g_m^2$ is small, the ``bare'' vortex mass ($\sim \gamma \sqrt Q$) is no longer subdominant to the electric field energy in \eqref{fieldEnergy}. Therefore, the contribution $\frac{1}{2^{1/4}} n\gamma \sqrt{Q}$ should be added to~\eqref{results2} and \eqref{results3},
where $n=2$ for $J\leq Q$, and $n= 3 J/Q$ for $J\gg Q$. For simplicity,  we assumed the same bare mass for all vortices. This gives the dominant spin dependent contribution to $\Delta$ for $J\lesssim Q/g_m^2$.

Consider now a weakly coupled electric theory. The essential difference in this case is that the cutoff is naturally identified with $\mu\sim g_e\sqrt{Q}$ \cite{alvarezGaume} instead of with~\eqref{cutoff}. Therefore, a single phonon is restricted to $J \lesssim g_e\sqrt{Q}$ and a vortex-antivortex pair to $J\gtrsim Q \mu^{-1}\sim\sqrt{Q}/g_e$. States with $J$ in the gap between the two consist of multiple phonons, approaching a $1/g_e^2$ number of them as $J \to \sqrt Q/g_e$. At this point, the lowest energy state shifts from multiple phonons to the vortex-antivortex pair. 
This consistently reflects the fact that vortices are now  heavy solitons and consist of also roughly a $1/g_e^2$ number of elementary  quanta. Since $\alpha \sim g_e$, the logarithmic term in~\eqref{results2} is indeed the expected result, $\mu/g_e^2\ln \mu d$, for a  semiclassical  solution with a vortex-antivortex pair split  by a distance $d=J/Q$. While possible a priori, we are not aware of any system in which the weakly coupled electric picture applies.

\paragraph*{Higher order corrections.}
On general grounds, we expect corrections to come from higher derivative terms controlled by the cutoff length scale $l \equiv \Lambda^{-1}$ \eqref{cutoff}. 
Two classes of effects are expected. The first class is controlled by the volume of the sphere and scales as $l^2 /R^2 \sim 1/Q$. The second class is controlled by the separation $d$ among vortices and scales as $l^2/d^2$, where the double power of~$d$ is dictated by rotational invariance. Using the relation $J\sim B d$ \eqref{J}, we have $l^2/d^2\sim Q/J^2$. 

In analyzing the possible terms in the Lagrangian, one indeed finds such corrections. The first class, already discussed in \cite{hellerman1,riccardo}, arises from higher derivative corrections paired by conformal invariance with terms suppressed by ${\cal R}/f$, where ${\cal R}$ is the Riemann tensor. The second class arises from the mass term in \eqref{action}. There we find relative corrections to the vortex action proportional to $\vec E^2/B^2$
and $\dot {\vec X} \wedge\vec E/ B$. These both scale as $l^2/d^2\sim Q/J^2$ on our solutions, becoming large at the lower edge $J\sim \sqrt Q$ of the double vortex states. 

Notice that for $\sqrt Q<J<Q$, the second class of corrections is larger than the $\O(1/J)$ quantum correction distinguishing~\eqref{quantum} from the the leading classical result~\eqref{results2}. Nonetheless, the quantum correction, even if numerically subleading, is functionally distinguished and thus calculable. We leave a systematic study of higher order corrections for future work.

\paragraph*{Discussion.}
To summarize, we calculated the scaling dimensions of operators with global charge $Q\gg 1$ and spin $J\lesssim Q^{3/2}$ by combining the state-operator correspondence with the EFT of vortices in superfluids. We also calculated correlation functions with two such operators and the Noether current. Other correlators as well as higher order corrections can be systematically computed. 


Our results apply to any CFT whose large charge sector is described by the EFT we presented. To be clear, we have not proved that there actually exists any such CFT. However, what are the possibilities? Given a state with finite charge density, the $U(1)$ symmetry may or may not be spontaneously broken. If it is broken, then the state is a superfluid and our results generically apply. If it is not broken, then the state is not a superfluid (e.g.,  a Fermi liquid) and our results do not apply. Because superfluids are such a natural possibility, we believe that there exists a large class of CFTs to which our results apply. Recently, Ref.~\cite{zhiboedov} framed this question within the conformal bootstrap.

Of course, it would be nice to explicitly identify such CFTs. One way forward is to consider CFTs that allow a perturbative expansion in some parameter and explicitly check if our results apply. For example, in $U(1)$ gauge theories, operators charged under the current \eqref{current} have been studied in a $1/N$ expansion \cite{largeN1,largeN2,largeN3,largeN4,largeN5,largeN6,largeN7,largeN8}, where~$N$ is the number of charged fields. The same operators were also studied in the $\epsilon$-expansion~\cite{epsilon}. In both cases, results were only given for small~$Q$, but the methods also apply at large~$Q$. Related to large $N$, large charge states have also been studied via the AdS/CFT correspondence under the name of ``holographic superconductors'' \cite{adscft1,adscft2,adscft3,adscft4,adscft5,adscft6,adscft7,adscft8}.

Perhaps AdS/CFT can also teach us how to study operators with $J \sim \Delta$ using EFT techniques, as this was the original motivation for the large spin bootstrap work  \cite{lightcone1,lightcone2,lightcone3,lightcone4,lightcone5,lightcone6,lightcone7,lightcone8,lightcone9,lightcone10,lightcone11}. The idea was that these operators should be described as widely separated---and therefore weakly interacting---objects in AdS space \cite{lightcone1}. This weak interaction suggests an EFT description, and such an EFT would then apply to all CFTs.

\begin{acknowledgments}
We acknowledge useful discussions with Joao Penedones.
The work of A.D., G.F.C., D.P. and R.R. is partially supported by the Swiss National Science Foundation  under contract 200020-169696 and through the National Center of Competence in Research SwissMAP. The work of A.M. was supported by the ERC-AdG-2015 grant 694896 and the Swiss National Science Foundation (Ambizione). D.P. is supported by European Union's Horizon 2020 Research Council grant 724659 MassiveCosmo ERC-2016-COG.

\end{acknowledgments}

\bibliography{references}

\begin{thebibliography}{60}%
\makeatletter
\providecommand \@ifxundefined [1]{%
 \@ifx{#1\undefined}
}%
\providecommand \@ifnum [1]{%
 \ifnum #1\expandafter \@firstoftwo
 \else \expandafter \@secondoftwo
 \fi
}%
\providecommand \@ifx [1]{%
 \ifx #1\expandafter \@firstoftwo
 \else \expandafter \@secondoftwo
 \fi
}%
\providecommand \natexlab [1]{#1}%
\providecommand \enquote  [1]{``#1''}%
\providecommand \bibnamefont  [1]{#1}%
\providecommand \bibfnamefont [1]{#1}%
\providecommand \citenamefont [1]{#1}%
\providecommand \href@noop [0]{\@secondoftwo}%
\providecommand \href [0]{\begingroup \@sanitize@url \@href}%
\providecommand \@href[1]{\@@startlink{#1}\@@href}%
\providecommand \@@href[1]{\endgroup#1\@@endlink}%
\providecommand \@sanitize@url [0]{\catcode `\\12\catcode `\$12\catcode
  `\&12\catcode `\#12\catcode `\^12\catcode `\_12\catcode `\%12\relax}%
\providecommand \@@startlink[1]{}%
\providecommand \@@endlink[0]{}%
\providecommand \url  [0]{\begingroup\@sanitize@url \@url }%
\providecommand \@url [1]{\endgroup\@href {#1}{\urlprefix }}%
\providecommand \urlprefix  [0]{URL }%
\providecommand \Eprint [0]{\href }%
\providecommand \doibase [0]{http://dx.doi.org/}%
\providecommand \selectlanguage [0]{\@gobble}%
\providecommand \bibinfo  [0]{\@secondoftwo}%
\providecommand \bibfield  [0]{\@secondoftwo}%
\providecommand \translation [1]{[#1]}%
\providecommand \BibitemOpen [0]{}%
\providecommand \bibitemStop [0]{}%
\providecommand \bibitemNoStop [0]{.\EOS\space}%
\providecommand \EOS [0]{\spacefactor3000\relax}%
\providecommand \BibitemShut  [1]{\csname bibitem#1\endcsname}%
\let\auto@bib@innerbib\@empty
\bibitem [{\citenamefont {Hellerman}\ \emph {et~al.}(2015)\citenamefont
  {Hellerman}, \citenamefont {Orlando}, \citenamefont {Reffert},\ and\
  \citenamefont {Watanabe}}]{hellerman1}%
  \BibitemOpen
  \bibfield  {author} {\bibinfo {author} {\bibfnamefont {S.}~\bibnamefont
  {Hellerman}}, \bibinfo {author} {\bibfnamefont {D.}~\bibnamefont {Orlando}},
  \bibinfo {author} {\bibfnamefont {S.}~\bibnamefont {Reffert}}, \ and\
  \bibinfo {author} {\bibfnamefont {M.}~\bibnamefont {Watanabe}},\ }\href
  {\doibase 10.1007/JHEP12(2015)071} {\bibfield  {journal} {\bibinfo  {journal}
  {JHEP}\ }\textbf {\bibinfo {volume} {12}},\ \bibinfo {pages} {071} (\bibinfo
  {year} {2015})},\ \Eprint {http://arxiv.org/abs/1505.01537} {arXiv:1505.01537
  [hep-th]} \BibitemShut {NoStop}%
\bibitem [{\citenamefont {Alvarez-Gaume}\ \emph {et~al.}(2017)\citenamefont
  {Alvarez-Gaume}, \citenamefont {Loukas}, \citenamefont {Orlando},\ and\
  \citenamefont {Reffert}}]{alvarezGaume}%
  \BibitemOpen
  \bibfield  {author} {\bibinfo {author} {\bibfnamefont {L.}~\bibnamefont
  {Alvarez-Gaume}}, \bibinfo {author} {\bibfnamefont {O.}~\bibnamefont
  {Loukas}}, \bibinfo {author} {\bibfnamefont {D.}~\bibnamefont {Orlando}}, \
  and\ \bibinfo {author} {\bibfnamefont {S.}~\bibnamefont {Reffert}},\ }\href
  {\doibase 10.1007/JHEP04(2017)059} {\bibfield  {journal} {\bibinfo  {journal}
  {JHEP}\ }\textbf {\bibinfo {volume} {04}},\ \bibinfo {pages} {059} (\bibinfo
  {year} {2017})},\ \Eprint {http://arxiv.org/abs/1610.04495} {arXiv:1610.04495
  [hep-th]} \BibitemShut {NoStop}%
\bibitem [{\citenamefont {Monin}\ \emph {et~al.}(2017)\citenamefont {Monin},
  \citenamefont {Pirtskhalava}, \citenamefont {Rattazzi},\ and\ \citenamefont
  {Seibold}}]{riccardo}%
  \BibitemOpen
  \bibfield  {author} {\bibinfo {author} {\bibfnamefont {A.}~\bibnamefont
  {Monin}}, \bibinfo {author} {\bibfnamefont {D.}~\bibnamefont {Pirtskhalava}},
  \bibinfo {author} {\bibfnamefont {R.}~\bibnamefont {Rattazzi}}, \ and\
  \bibinfo {author} {\bibfnamefont {F.~K.}\ \bibnamefont {Seibold}},\ }\href
  {\doibase 10.1007/JHEP06(2017)011} {\bibfield  {journal} {\bibinfo  {journal}
  {JHEP}\ }\textbf {\bibinfo {volume} {06}},\ \bibinfo {pages} {011} (\bibinfo
  {year} {2017})},\ \Eprint {http://arxiv.org/abs/1611.02912} {arXiv:1611.02912
  [hep-th]} \BibitemShut {NoStop}%
\bibitem [{\citenamefont {Loukas}(2016)}]{loukas}%
  \BibitemOpen
  \bibfield  {author} {\bibinfo {author} {\bibfnamefont {O.}~\bibnamefont
  {Loukas}},\ }\href@noop {} {\  (\bibinfo {year} {2016})},\ \Eprint
  {http://arxiv.org/abs/1612.08985} {arXiv:1612.08985 [hep-th]} \BibitemShut
  {NoStop}%
\bibitem [{\citenamefont {Hellerman}\ \emph
  {et~al.}(2017{\natexlab{a}})\citenamefont {Hellerman}, \citenamefont
  {Kobayashi}, \citenamefont {Maeda},\ and\ \citenamefont
  {Watanabe}}]{hellerman2}%
  \BibitemOpen
  \bibfield  {author} {\bibinfo {author} {\bibfnamefont {S.}~\bibnamefont
  {Hellerman}}, \bibinfo {author} {\bibfnamefont {N.}~\bibnamefont
  {Kobayashi}}, \bibinfo {author} {\bibfnamefont {S.}~\bibnamefont {Maeda}}, \
  and\ \bibinfo {author} {\bibfnamefont {M.}~\bibnamefont {Watanabe}},\
  }\href@noop {} {\  (\bibinfo {year} {2017}{\natexlab{a}})},\ \Eprint
  {http://arxiv.org/abs/1705.05825} {arXiv:1705.05825 [hep-th]} \BibitemShut
  {NoStop}%
\bibitem [{\citenamefont {Hellerman}\ \emph
  {et~al.}(2017{\natexlab{b}})\citenamefont {Hellerman}, \citenamefont
  {Maeda},\ and\ \citenamefont {Watanabe}}]{hellerman3}%
  \BibitemOpen
  \bibfield  {author} {\bibinfo {author} {\bibfnamefont {S.}~\bibnamefont
  {Hellerman}}, \bibinfo {author} {\bibfnamefont {S.}~\bibnamefont {Maeda}}, \
  and\ \bibinfo {author} {\bibfnamefont {M.}~\bibnamefont {Watanabe}},\
  }\href@noop {} {\  (\bibinfo {year} {2017}{\natexlab{b}})},\ \Eprint
  {http://arxiv.org/abs/1706.05743} {arXiv:1706.05743 [hep-th]} \BibitemShut
  {NoStop}%
\bibitem [{\citenamefont {Banerjee}\ \emph {et~al.}(2017)\citenamefont
  {Banerjee}, \citenamefont {Chandrasekharan},\ and\ \citenamefont
  {Orlando}}]{monteCarlo}%
  \BibitemOpen
  \bibfield  {author} {\bibinfo {author} {\bibfnamefont {D.}~\bibnamefont
  {Banerjee}}, \bibinfo {author} {\bibfnamefont {S.}~\bibnamefont
  {Chandrasekharan}}, \ and\ \bibinfo {author} {\bibfnamefont {D.}~\bibnamefont
  {Orlando}},\ }\href@noop {} {\  (\bibinfo {year} {2017})},\ \Eprint
  {http://arxiv.org/abs/1707.00711} {arXiv:1707.00711 [hep-lat]} \BibitemShut
  {NoStop}%
\bibitem [{\citenamefont {Loukas}\ \emph {et~al.}(2017)\citenamefont {Loukas},
  \citenamefont {Orlando},\ and\ \citenamefont {Reffert}}]{matrixModels}%
  \BibitemOpen
  \bibfield  {author} {\bibinfo {author} {\bibfnamefont {O.}~\bibnamefont
  {Loukas}}, \bibinfo {author} {\bibfnamefont {D.}~\bibnamefont {Orlando}}, \
  and\ \bibinfo {author} {\bibfnamefont {S.}~\bibnamefont {Reffert}},\
  }\href@noop {} {\  (\bibinfo {year} {2017})},\ \Eprint
  {http://arxiv.org/abs/1707.00710} {arXiv:1707.00710 [hep-th]} \BibitemShut
  {NoStop}%
\bibitem [{\citenamefont {Hellerman}\ and\ \citenamefont
  {Maeda}(2017)}]{hellerman4}%
  \BibitemOpen
  \bibfield  {author} {\bibinfo {author} {\bibfnamefont {S.}~\bibnamefont
  {Hellerman}}\ and\ \bibinfo {author} {\bibfnamefont {S.}~\bibnamefont
  {Maeda}},\ }\href@noop {} {\  (\bibinfo {year} {2017})},\ \Eprint
  {http://arxiv.org/abs/1710.07336} {arXiv:1710.07336 [hep-th]} \BibitemShut
  {NoStop}%
\bibitem [{\citenamefont {Son}(2002)}]{son}%
  \BibitemOpen
  \bibfield  {author} {\bibinfo {author} {\bibfnamefont {D.~T.}\ \bibnamefont
  {Son}},\ }\href@noop {} {\  (\bibinfo {year} {2002})},\ \Eprint
  {http://arxiv.org/abs/hep-ph/0204199} {arXiv:hep-ph/0204199 [hep-ph]}
  \BibitemShut {NoStop}%
\bibitem [{\citenamefont {Rychkov}(2016)}]{slavaCFT}%
  \BibitemOpen
  \bibfield  {author} {\bibinfo {author} {\bibfnamefont {S.}~\bibnamefont
  {Rychkov}},\ }\href {\doibase 10.1007/978-3-319-43626-5} {\emph {\bibinfo
  {title} {{EPFL Lectures on Conformal Field Theory in D>= 3 Dimensions}}}},\
  SpringerBriefs in Physics\ (\bibinfo {year} {2016})\ \Eprint
  {http://arxiv.org/abs/1601.05000} {arXiv:1601.05000 [hep-th]} \BibitemShut
  {NoStop}%
\bibitem [{\citenamefont {{Simmons-Duffin, D.}}(2016)}]{dsdCFT}%
  \BibitemOpen
  \bibfield  {author} {\bibinfo {author} {\bibnamefont {{Simmons-Duffin,
  D.}}},\ }\enquote {\bibinfo {title} {{The Conformal Bootstrap}},}\ in\ \href
  {\doibase {10.1142/9789813149441_0001}} {\emph {\bibinfo {booktitle} {{TASI
  2015: New Frontiers in Fields and Strings}}}}\ (\bibinfo  {publisher} {World
  Scientific},\ \bibinfo {year} {2016})\ Chap.~\bibinfo {chapter} {{1}}, pp.\
  \bibinfo {pages} {{1--74}},\ \Eprint {http://arxiv.org/abs/1602.07982}
  {arXiv:1602.07982 [hep-th]} \BibitemShut {NoStop}%
\bibitem [{\citenamefont {Fitzpatrick}\ \emph {et~al.}(2013)\citenamefont
  {Fitzpatrick}, \citenamefont {Kaplan}, \citenamefont {Poland},\ and\
  \citenamefont {Simmons-Duffin}}]{lightcone1}%
  \BibitemOpen
  \bibfield  {author} {\bibinfo {author} {\bibfnamefont {A.~L.}\ \bibnamefont
  {Fitzpatrick}}, \bibinfo {author} {\bibfnamefont {J.}~\bibnamefont {Kaplan}},
  \bibinfo {author} {\bibfnamefont {D.}~\bibnamefont {Poland}}, \ and\ \bibinfo
  {author} {\bibfnamefont {D.}~\bibnamefont {Simmons-Duffin}},\ }\href
  {\doibase 10.1007/JHEP12(2013)004} {\bibfield  {journal} {\bibinfo  {journal}
  {JHEP}\ }\textbf {\bibinfo {volume} {12}},\ \bibinfo {pages} {004} (\bibinfo
  {year} {2013})},\ \Eprint {http://arxiv.org/abs/1212.3616} {arXiv:1212.3616
  [hep-th]} \BibitemShut {NoStop}%
\bibitem [{\citenamefont {Komargodski}\ and\ \citenamefont
  {Zhiboedov}(2013)}]{lightcone2}%
  \BibitemOpen
  \bibfield  {author} {\bibinfo {author} {\bibfnamefont {Z.}~\bibnamefont
  {Komargodski}}\ and\ \bibinfo {author} {\bibfnamefont {A.}~\bibnamefont
  {Zhiboedov}},\ }\href {\doibase 10.1007/JHEP11(2013)140} {\bibfield
  {journal} {\bibinfo  {journal} {JHEP}\ }\textbf {\bibinfo {volume} {11}},\
  \bibinfo {pages} {140} (\bibinfo {year} {2013})},\ \Eprint
  {http://arxiv.org/abs/1212.4103} {arXiv:1212.4103 [hep-th]} \BibitemShut
  {NoStop}%
\bibitem [{\citenamefont {Kaviraj}\ \emph
  {et~al.}(2015{\natexlab{a}})\citenamefont {Kaviraj}, \citenamefont {Sen},\
  and\ \citenamefont {Sinha}}]{lightcone3}%
  \BibitemOpen
  \bibfield  {author} {\bibinfo {author} {\bibfnamefont {A.}~\bibnamefont
  {Kaviraj}}, \bibinfo {author} {\bibfnamefont {K.}~\bibnamefont {Sen}}, \ and\
  \bibinfo {author} {\bibfnamefont {A.}~\bibnamefont {Sinha}},\ }\href
  {\doibase 10.1007/JHEP11(2015)083} {\bibfield  {journal} {\bibinfo  {journal}
  {JHEP}\ }\textbf {\bibinfo {volume} {11}},\ \bibinfo {pages} {083} (\bibinfo
  {year} {2015}{\natexlab{a}})},\ \Eprint {http://arxiv.org/abs/1502.01437}
  {arXiv:1502.01437 [hep-th]} \BibitemShut {NoStop}%
\bibitem [{\citenamefont {Alday}\ \emph {et~al.}(2015)\citenamefont {Alday},
  \citenamefont {Bissi},\ and\ \citenamefont {Lukowski}}]{lightcone4}%
  \BibitemOpen
  \bibfield  {author} {\bibinfo {author} {\bibfnamefont {L.~F.}\ \bibnamefont
  {Alday}}, \bibinfo {author} {\bibfnamefont {A.}~\bibnamefont {Bissi}}, \ and\
  \bibinfo {author} {\bibfnamefont {T.}~\bibnamefont {Lukowski}},\ }\href
  {\doibase 10.1007/JHEP11(2015)101} {\bibfield  {journal} {\bibinfo  {journal}
  {JHEP}\ }\textbf {\bibinfo {volume} {11}},\ \bibinfo {pages} {101} (\bibinfo
  {year} {2015})},\ \Eprint {http://arxiv.org/abs/1502.07707} {arXiv:1502.07707
  [hep-th]} \BibitemShut {NoStop}%
\bibitem [{\citenamefont {Kaviraj}\ \emph
  {et~al.}(2015{\natexlab{b}})\citenamefont {Kaviraj}, \citenamefont {Sen},\
  and\ \citenamefont {Sinha}}]{lightcone5}%
  \BibitemOpen
  \bibfield  {author} {\bibinfo {author} {\bibfnamefont {A.}~\bibnamefont
  {Kaviraj}}, \bibinfo {author} {\bibfnamefont {K.}~\bibnamefont {Sen}}, \ and\
  \bibinfo {author} {\bibfnamefont {A.}~\bibnamefont {Sinha}},\ }\href
  {\doibase 10.1007/JHEP07(2015)026} {\bibfield  {journal} {\bibinfo  {journal}
  {JHEP}\ }\textbf {\bibinfo {volume} {07}},\ \bibinfo {pages} {026} (\bibinfo
  {year} {2015}{\natexlab{b}})},\ \Eprint {http://arxiv.org/abs/1504.00772}
  {arXiv:1504.00772 [hep-th]} \BibitemShut {NoStop}%
\bibitem [{\citenamefont {Alday}\ and\ \citenamefont
  {Zhiboedov}(2016)}]{lightcone6}%
  \BibitemOpen
  \bibfield  {author} {\bibinfo {author} {\bibfnamefont {L.~F.}\ \bibnamefont
  {Alday}}\ and\ \bibinfo {author} {\bibfnamefont {A.}~\bibnamefont
  {Zhiboedov}},\ }\href {\doibase 10.1007/JHEP06(2016)091} {\bibfield
  {journal} {\bibinfo  {journal} {JHEP}\ }\textbf {\bibinfo {volume} {06}},\
  \bibinfo {pages} {091} (\bibinfo {year} {2016})},\ \Eprint
  {http://arxiv.org/abs/1506.04659} {arXiv:1506.04659 [hep-th]} \BibitemShut
  {NoStop}%
\bibitem [{\citenamefont {Alday}\ and\ \citenamefont
  {Zhiboedov}(2017)}]{lightcone7}%
  \BibitemOpen
  \bibfield  {author} {\bibinfo {author} {\bibfnamefont {L.~F.}\ \bibnamefont
  {Alday}}\ and\ \bibinfo {author} {\bibfnamefont {A.}~\bibnamefont
  {Zhiboedov}},\ }\href {\doibase 10.1007/JHEP04(2017)157} {\bibfield
  {journal} {\bibinfo  {journal} {JHEP}\ }\textbf {\bibinfo {volume} {04}},\
  \bibinfo {pages} {157} (\bibinfo {year} {2017})},\ \Eprint
  {http://arxiv.org/abs/1510.08091} {arXiv:1510.08091 [hep-th]} \BibitemShut
  {NoStop}%
\bibitem [{\citenamefont {Li}\ \emph {et~al.}(2016)\citenamefont {Li},
  \citenamefont {Meltzer},\ and\ \citenamefont {Poland}}]{lightcone8}%
  \BibitemOpen
  \bibfield  {author} {\bibinfo {author} {\bibfnamefont {D.}~\bibnamefont
  {Li}}, \bibinfo {author} {\bibfnamefont {D.}~\bibnamefont {Meltzer}}, \ and\
  \bibinfo {author} {\bibfnamefont {D.}~\bibnamefont {Poland}},\ }\href
  {\doibase 10.1007/JHEP02(2016)143} {\bibfield  {journal} {\bibinfo  {journal}
  {JHEP}\ }\textbf {\bibinfo {volume} {02}},\ \bibinfo {pages} {143} (\bibinfo
  {year} {2016})},\ \Eprint {http://arxiv.org/abs/1511.08025} {arXiv:1511.08025
  [hep-th]} \BibitemShut {NoStop}%
\bibitem [{\citenamefont {Simmons-Duffin}(2017)}]{lightcone9}%
  \BibitemOpen
  \bibfield  {author} {\bibinfo {author} {\bibfnamefont {D.}~\bibnamefont
  {Simmons-Duffin}},\ }\href {\doibase 10.1007/JHEP03(2017)086} {\bibfield
  {journal} {\bibinfo  {journal} {JHEP}\ }\textbf {\bibinfo {volume} {03}},\
  \bibinfo {pages} {086} (\bibinfo {year} {2017})},\ \Eprint
  {http://arxiv.org/abs/1612.08471} {arXiv:1612.08471 [hep-th]} \BibitemShut
  {NoStop}%
\bibitem [{\citenamefont {Caron-Huot}(2017)}]{lightcone10}%
  \BibitemOpen
  \bibfield  {author} {\bibinfo {author} {\bibfnamefont {S.}~\bibnamefont
  {Caron-Huot}},\ }\href {\doibase 10.1007/JHEP09(2017)078} {\bibfield
  {journal} {\bibinfo  {journal} {JHEP}\ }\textbf {\bibinfo {volume} {09}},\
  \bibinfo {pages} {078} (\bibinfo {year} {2017})},\ \Eprint
  {http://arxiv.org/abs/1703.00278} {arXiv:1703.00278 [hep-th]} \BibitemShut
  {NoStop}%
\bibitem [{\citenamefont {Qiao}\ and\ \citenamefont
  {Rychkov}(2017)}]{lightcone11}%
  \BibitemOpen
  \bibfield  {author} {\bibinfo {author} {\bibfnamefont {J.}~\bibnamefont
  {Qiao}}\ and\ \bibinfo {author} {\bibfnamefont {S.}~\bibnamefont {Rychkov}},\
  }\href@noop {} {\  (\bibinfo {year} {2017})},\ \Eprint
  {http://arxiv.org/abs/1709.00008} {arXiv:1709.00008 [hep-th]} \BibitemShut
  {NoStop}%
\bibitem [{\citenamefont {Donnelly}(1991)}]{vortexBook}%
  \BibitemOpen
  \bibfield  {author} {\bibinfo {author} {\bibfnamefont {R.}~\bibnamefont
  {Donnelly}},\ }\href
  {http://www.cambridge.org/catalogue/catalogue.asp?isbn=9780521018142} {\emph
  {\bibinfo {title} {Quantized Vortices in Helium II}}},\ \bibinfo {series}
  {Cambridge Studies in Low Temperature Physics}\ No.~\bibinfo {number} {3}\
  (\bibinfo  {publisher} {Cambridge University Press},\ \bibinfo {year}
  {1991})\BibitemShut {NoStop}%
\bibitem [{\citenamefont {Horn}\ \emph {et~al.}(2015)\citenamefont {Horn},
  \citenamefont {Nicolis},\ and\ \citenamefont {Penco}}]{nicolis}%
  \BibitemOpen
  \bibfield  {author} {\bibinfo {author} {\bibfnamefont {B.}~\bibnamefont
  {Horn}}, \bibinfo {author} {\bibfnamefont {A.}~\bibnamefont {Nicolis}}, \
  and\ \bibinfo {author} {\bibfnamefont {R.}~\bibnamefont {Penco}},\ }\href
  {\doibase 10.1007/JHEP10(2015)153} {\bibfield  {journal} {\bibinfo  {journal}
  {JHEP}\ }\textbf {\bibinfo {volume} {10}},\ \bibinfo {pages} {153} (\bibinfo
  {year} {2015})},\ \Eprint {http://arxiv.org/abs/1507.05635} {arXiv:1507.05635
  [hep-th]} \BibitemShut {NoStop}%
\bibitem [{\citenamefont {Feynman}(1998)}]{feynman}%
  \BibitemOpen
  \bibfield  {author} {\bibinfo {author} {\bibfnamefont {R.}~\bibnamefont
  {Feynman}},\ }\href {https://books.google.ch/books?id=Ou4ltPYiXPgC} {\emph
  {\bibinfo {title} {Statistical Mechanics: A Set Of Lectures}}},\ Advanced
  Books Classics\ (\bibinfo  {publisher} {Avalon Publishing},\ \bibinfo {year}
  {1998})\BibitemShut {NoStop}%
\bibitem [{\citenamefont {Wilson}\ and\ \citenamefont
  {Fisher}(1972)}]{wilsonFisher}%
  \BibitemOpen
  \bibfield  {author} {\bibinfo {author} {\bibfnamefont {K.~G.}\ \bibnamefont
  {Wilson}}\ and\ \bibinfo {author} {\bibfnamefont {M.~E.}\ \bibnamefont
  {Fisher}},\ }\href {\doibase 10.1103/PhysRevLett.28.240} {\bibfield
  {journal} {\bibinfo  {journal} {Phys. Rev. Lett.}\ }\textbf {\bibinfo
  {volume} {28}},\ \bibinfo {pages} {240} (\bibinfo {year} {1972})}\BibitemShut
  {NoStop}%
\bibitem [{\citenamefont {Zee}(2010)}]{zee}%
  \BibitemOpen
  \bibfield  {author} {\bibinfo {author} {\bibfnamefont {A.}~\bibnamefont
  {Zee}},\ }\href
  {https://www.kitp.ucsb.edu/zee/books/quantum-field-theory-nutshell} {\emph
  {\bibinfo {title} {Quantum Field Theory in a Nutshell: Second Edition}}},\ In
  a Nutshell\ (\bibinfo  {publisher} {Princeton University Press},\ \bibinfo
  {year} {2010})\BibitemShut {NoStop}%
\bibitem [{\citenamefont {Landau}\ and\ \citenamefont
  {Lifshitz}(1975)}]{landauCTF}%
  \BibitemOpen
  \bibfield  {author} {\bibinfo {author} {\bibfnamefont {L.}~\bibnamefont
  {Landau}}\ and\ \bibinfo {author} {\bibfnamefont {E.}~\bibnamefont
  {Lifshitz}},\ }\href@noop {} {\emph {\bibinfo {title} {The Classical Theory
  of Fields}}},\ \bibinfo {edition} {4th}\ ed.,\ Course of Theoretical Physics\
  (\bibinfo  {publisher} {Butterworth-Heinemann},\ \bibinfo {year}
  {1975})\BibitemShut {NoStop}%
\bibitem [{\citenamefont {Landau}\ and\ \citenamefont
  {Lifshitz}(1977)}]{landauQM}%
  \BibitemOpen
  \bibfield  {author} {\bibinfo {author} {\bibfnamefont {L.}~\bibnamefont
  {Landau}}\ and\ \bibinfo {author} {\bibfnamefont {E.}~\bibnamefont
  {Lifshitz}},\ }\href@noop {} {\emph {\bibinfo {title} {Quantum Mechanics}}},\
  \bibinfo {edition} {3rd}\ ed.,\ Course of Theoretical Physics\ (\bibinfo
  {publisher} {Butterworth-Heinemann},\ \bibinfo {year} {1977})\BibitemShut
  {NoStop}%
\bibitem [{\citenamefont {Sivan}\ and\ \citenamefont
  {Levit}(1992)}]{integrateLandau1}%
  \BibitemOpen
  \bibfield  {author} {\bibinfo {author} {\bibfnamefont {N.}~\bibnamefont
  {Sivan}}\ and\ \bibinfo {author} {\bibfnamefont {S.}~\bibnamefont {Levit}},\
  }\href {\doibase 10.1103/PhysRevB.46.2319} {\bibfield  {journal} {\bibinfo
  {journal} {Phys. Rev. B}\ }\textbf {\bibinfo {volume} {46}},\ \bibinfo
  {pages} {2319} (\bibinfo {year} {1992})}\BibitemShut {NoStop}%
\bibitem [{\citenamefont {Entelis}\ and\ \citenamefont
  {Levit}(1992)}]{integrateLandau2}%
  \BibitemOpen
  \bibfield  {author} {\bibinfo {author} {\bibfnamefont {A.}~\bibnamefont
  {Entelis}}\ and\ \bibinfo {author} {\bibfnamefont {S.}~\bibnamefont
  {Levit}},\ }\href {\doibase 10.1103/PhysRevLett.69.3001} {\bibfield
  {journal} {\bibinfo  {journal} {Phys. Rev. Lett.}\ }\textbf {\bibinfo
  {volume} {69}},\ \bibinfo {pages} {3001} (\bibinfo {year}
  {1992})}\BibitemShut {NoStop}%
\bibitem [{\citenamefont {Tochishita}\ \emph {et~al.}(1996)\citenamefont
  {Tochishita}, \citenamefont {Mizui},\ and\ \citenamefont
  {Kuratsuji}}]{integrateLandau3}%
  \BibitemOpen
  \bibfield  {author} {\bibinfo {author} {\bibfnamefont {T.}~\bibnamefont
  {Tochishita}}, \bibinfo {author} {\bibfnamefont {M.}~\bibnamefont {Mizui}}, \
  and\ \bibinfo {author} {\bibfnamefont {H.}~\bibnamefont {Kuratsuji}},\ }\href
  {\doibase 10.1016/0375-9601(96)00073-4} {\bibfield  {journal} {\bibinfo
  {journal} {Physics Letters A}\ }\textbf {\bibinfo {volume} {212}},\ \bibinfo
  {pages} {304 } (\bibinfo {year} {1996})}\BibitemShut {NoStop}%
\bibitem [{\citenamefont {Dunne}\ \emph {et~al.}(1990)\citenamefont {Dunne},
  \citenamefont {Jackiw},\ and\ \citenamefont {Trugenberger}}]{jackiw1}%
  \BibitemOpen
  \bibfield  {author} {\bibinfo {author} {\bibfnamefont {G.~V.}\ \bibnamefont
  {Dunne}}, \bibinfo {author} {\bibfnamefont {R.}~\bibnamefont {Jackiw}}, \
  and\ \bibinfo {author} {\bibfnamefont {C.~A.}\ \bibnamefont {Trugenberger}},\
  }\href {\doibase 10.1103/PhysRevD.41.661} {\bibfield  {journal} {\bibinfo
  {journal} {Phys. Rev. D}\ }\textbf {\bibinfo {volume} {41}},\ \bibinfo
  {pages} {661} (\bibinfo {year} {1990})}\BibitemShut {NoStop}%
\bibitem [{\citenamefont {Dunne}\ and\ \citenamefont {Jackiw}(1993)}]{jackiw2}%
  \BibitemOpen
  \bibfield  {author} {\bibinfo {author} {\bibfnamefont {G.}~\bibnamefont
  {Dunne}}\ and\ \bibinfo {author} {\bibfnamefont {R.}~\bibnamefont {Jackiw}},\
  }\href {\doibase 10.1016/0920-5632(93)90376-H} {\bibfield  {journal}
  {\bibinfo  {journal} {Nuclear Physics B - Proceedings Supplements}\ }\textbf
  {\bibinfo {volume} {33}},\ \bibinfo {pages} {114 } (\bibinfo {year}
  {1993})}\BibitemShut {NoStop}%
\bibitem [{\citenamefont {Wu}\ and\ \citenamefont
  {Yang}(1976)}]{monopoleHarmonics}%
  \BibitemOpen
  \bibfield  {author} {\bibinfo {author} {\bibfnamefont {T.~T.}\ \bibnamefont
  {Wu}}\ and\ \bibinfo {author} {\bibfnamefont {C.~N.}\ \bibnamefont {Yang}},\
  }\href {\doibase 10.1016/0550-3213(76)90143-7} {\bibfield  {journal}
  {\bibinfo  {journal} {Nuclear Physics B}\ }\textbf {\bibinfo {volume}
  {107}},\ \bibinfo {pages} {365 } (\bibinfo {year} {1976})}\BibitemShut
  {NoStop}%
\bibitem [{\citenamefont {Coleman}(1983)}]{coleman}%
  \BibitemOpen
  \bibfield  {author} {\bibinfo {author} {\bibfnamefont {S.}~\bibnamefont
  {Coleman}},\ }\enquote {\bibinfo {title} {The magnetic monopole fifty years
  later},}\ in\ \href {\doibase 10.1007/978-1-4613-3655-6_2} {\emph {\bibinfo
  {booktitle} {The Unity of the Fundamental Interactions}}},\ \bibinfo {editor}
  {edited by\ \bibinfo {editor} {\bibfnamefont {A.}~\bibnamefont {Zichichi}}}\
  (\bibinfo  {publisher} {Springer US},\ \bibinfo {address} {Boston, MA},\
  \bibinfo {year} {1983})\ pp.\ \bibinfo {pages} {21--117}\BibitemShut
  {NoStop}%
\bibitem [{\citenamefont {Hatsuda}\ \emph {et~al.}(2003)\citenamefont
  {Hatsuda}, \citenamefont {Iso},\ and\ \citenamefont {Umetsu}}]{fuzzySphere1}%
  \BibitemOpen
  \bibfield  {author} {\bibinfo {author} {\bibfnamefont {M.}~\bibnamefont
  {Hatsuda}}, \bibinfo {author} {\bibfnamefont {S.}~\bibnamefont {Iso}}, \ and\
  \bibinfo {author} {\bibfnamefont {H.}~\bibnamefont {Umetsu}},\ }\href
  {\doibase 10.1016/j.nuclphysb.2003.08.013} {\bibfield  {journal} {\bibinfo
  {journal} {Nuclear Physics B}\ }\textbf {\bibinfo {volume} {671}},\ \bibinfo
  {pages} {217 } (\bibinfo {year} {2003})},\ \Eprint
  {http://arxiv.org/abs/hep-th/0306251} {arXiv:hep-th/0306251 [hep-th]}
  \BibitemShut {NoStop}%
\bibitem [{\citenamefont {Hasebe}(2010)}]{fuzzySphere2}%
  \BibitemOpen
  \bibfield  {author} {\bibinfo {author} {\bibfnamefont {K.}~\bibnamefont
  {Hasebe}},\ }\href {\doibase 10.3842/SIGMA.2010.071} {\bibfield  {journal}
  {\bibinfo  {journal} {SIGMA}\ }\textbf {\bibinfo {volume} {6}},\ \bibinfo
  {pages} {071} (\bibinfo {year} {2010})},\ \Eprint
  {http://arxiv.org/abs/1009.1192} {arXiv:1009.1192 [hep-th]} \BibitemShut
  {NoStop}%
\bibitem [{\citenamefont {Haldane}(1983)}]{haldane}%
  \BibitemOpen
  \bibfield  {author} {\bibinfo {author} {\bibfnamefont {F.~D.~M.}\
  \bibnamefont {Haldane}},\ }\href {\doibase 10.1103/PhysRevLett.51.605}
  {\bibfield  {journal} {\bibinfo  {journal} {Phys. Rev. Lett.}\ }\textbf
  {\bibinfo {volume} {51}},\ \bibinfo {pages} {605} (\bibinfo {year}
  {1983})}\BibitemShut {NoStop}%
\bibitem [{\citenamefont {Greiter}(2011)}]{greiter}%
  \BibitemOpen
  \bibfield  {author} {\bibinfo {author} {\bibfnamefont {M.}~\bibnamefont
  {Greiter}},\ }\href {\doibase 10.1103/PhysRevB.83.115129} {\bibfield
  {journal} {\bibinfo  {journal} {Phys. Rev. B}\ }\textbf {\bibinfo {volume}
  {83}},\ \bibinfo {pages} {115129} (\bibinfo {year} {2011})},\ \Eprint
  {http://arxiv.org/abs/1101.3943} {arXiv:1101.3943 [cond-mat]} \BibitemShut
  {NoStop}%
\bibitem [{\citenamefont {Costa}\ \emph {et~al.}(2011)\citenamefont {Costa},
  \citenamefont {Penedones}, \citenamefont {Poland},\ and\ \citenamefont
  {Rychkov}}]{spinning}%
  \BibitemOpen
  \bibfield  {author} {\bibinfo {author} {\bibfnamefont {M.~S.}\ \bibnamefont
  {Costa}}, \bibinfo {author} {\bibfnamefont {J.}~\bibnamefont {Penedones}},
  \bibinfo {author} {\bibfnamefont {D.}~\bibnamefont {Poland}}, \ and\ \bibinfo
  {author} {\bibfnamefont {S.}~\bibnamefont {Rychkov}},\ }\href {\doibase
  10.1007/JHEP11(2011)071} {\bibfield  {journal} {\bibinfo  {journal} {JHEP}\
  }\textbf {\bibinfo {volume} {11}},\ \bibinfo {pages} {071} (\bibinfo {year}
  {2011})},\ \Eprint {http://arxiv.org/abs/1107.3554} {arXiv:1107.3554
  [hep-th]} \BibitemShut {NoStop}%
\bibitem [{\citenamefont {Murthy}\ and\ \citenamefont
  {Sachdev}(1990)}]{largeN1}%
  \BibitemOpen
  \bibfield  {author} {\bibinfo {author} {\bibfnamefont {G.}~\bibnamefont
  {Murthy}}\ and\ \bibinfo {author} {\bibfnamefont {S.}~\bibnamefont
  {Sachdev}},\ }\href {\doibase 10.1016/0550-3213(90)90670-9} {\bibfield
  {journal} {\bibinfo  {journal} {Nucl. Phys.}\ }\textbf {\bibinfo {volume}
  {B344}},\ \bibinfo {pages} {557} (\bibinfo {year} {1990})}\BibitemShut
  {NoStop}%
\bibitem [{\citenamefont {Borokhov}\ \emph {et~al.}(2002)\citenamefont
  {Borokhov}, \citenamefont {Kapustin},\ and\ \citenamefont {Wu}}]{largeN2}%
  \BibitemOpen
  \bibfield  {author} {\bibinfo {author} {\bibfnamefont {V.}~\bibnamefont
  {Borokhov}}, \bibinfo {author} {\bibfnamefont {A.}~\bibnamefont {Kapustin}},
  \ and\ \bibinfo {author} {\bibfnamefont {X.-k.}\ \bibnamefont {Wu}},\ }\href
  {\doibase 10.1088/1126-6708/2002/11/049} {\bibfield  {journal} {\bibinfo
  {journal} {JHEP}\ }\textbf {\bibinfo {volume} {11}},\ \bibinfo {pages} {049}
  (\bibinfo {year} {2002})},\ \Eprint {http://arxiv.org/abs/hep-th/0206054}
  {arXiv:hep-th/0206054 [hep-th]} \BibitemShut {NoStop}%
\bibitem [{\citenamefont {Metlitski}\ \emph {et~al.}(2008)\citenamefont
  {Metlitski}, \citenamefont {Hermele}, \citenamefont {Senthil},\ and\
  \citenamefont {Fisher}}]{largeN3}%
  \BibitemOpen
  \bibfield  {author} {\bibinfo {author} {\bibfnamefont {M.~A.}\ \bibnamefont
  {Metlitski}}, \bibinfo {author} {\bibfnamefont {M.}~\bibnamefont {Hermele}},
  \bibinfo {author} {\bibfnamefont {T.}~\bibnamefont {Senthil}}, \ and\
  \bibinfo {author} {\bibfnamefont {M.~P.~A.}\ \bibnamefont {Fisher}},\ }\href
  {\doibase 10.1103/PhysRevB.78.214418} {\bibfield  {journal} {\bibinfo
  {journal} {Phys. Rev.}\ }\textbf {\bibinfo {volume} {B78}},\ \bibinfo {pages}
  {214418} (\bibinfo {year} {2008})},\ \Eprint {http://arxiv.org/abs/0809.2816}
  {arXiv:0809.2816 [cond-mat.str-el]} \BibitemShut {NoStop}%
\bibitem [{\citenamefont {Pufu}\ and\ \citenamefont {Sachdev}(2013)}]{largeN4}%
  \BibitemOpen
  \bibfield  {author} {\bibinfo {author} {\bibfnamefont {S.~S.}\ \bibnamefont
  {Pufu}}\ and\ \bibinfo {author} {\bibfnamefont {S.}~\bibnamefont {Sachdev}},\
  }\href {\doibase 10.1007/JHEP09(2013)127} {\bibfield  {journal} {\bibinfo
  {journal} {JHEP}\ }\textbf {\bibinfo {volume} {09}},\ \bibinfo {pages} {127}
  (\bibinfo {year} {2013})},\ \Eprint {http://arxiv.org/abs/1303.3006}
  {arXiv:1303.3006 [hep-th]} \BibitemShut {NoStop}%
\bibitem [{\citenamefont {Pufu}(2014)}]{largeN5}%
  \BibitemOpen
  \bibfield  {author} {\bibinfo {author} {\bibfnamefont {S.~S.}\ \bibnamefont
  {Pufu}},\ }\href {\doibase 10.1103/PhysRevD.89.065016} {\bibfield  {journal}
  {\bibinfo  {journal} {Phys. Rev.}\ }\textbf {\bibinfo {volume} {D89}},\
  \bibinfo {pages} {065016} (\bibinfo {year} {2014})},\ \Eprint
  {http://arxiv.org/abs/1303.6125} {arXiv:1303.6125 [hep-th]} \BibitemShut
  {NoStop}%
\bibitem [{\citenamefont {Dyer}\ \emph {et~al.}(2013)\citenamefont {Dyer},
  \citenamefont {Mezei},\ and\ \citenamefont {Pufu}}]{largeN6}%
  \BibitemOpen
  \bibfield  {author} {\bibinfo {author} {\bibfnamefont {E.}~\bibnamefont
  {Dyer}}, \bibinfo {author} {\bibfnamefont {M.}~\bibnamefont {Mezei}}, \ and\
  \bibinfo {author} {\bibfnamefont {S.~S.}\ \bibnamefont {Pufu}},\ }\href@noop
  {} {\  (\bibinfo {year} {2013})},\ \Eprint {http://arxiv.org/abs/1309.1160}
  {arXiv:1309.1160 [hep-th]} \BibitemShut {NoStop}%
\bibitem [{\citenamefont {Dyer}\ \emph {et~al.}(2015)\citenamefont {Dyer},
  \citenamefont {Mezei}, \citenamefont {Pufu},\ and\ \citenamefont
  {Sachdev}}]{largeN7}%
  \BibitemOpen
  \bibfield  {author} {\bibinfo {author} {\bibfnamefont {E.}~\bibnamefont
  {Dyer}}, \bibinfo {author} {\bibfnamefont {M.}~\bibnamefont {Mezei}},
  \bibinfo {author} {\bibfnamefont {S.~S.}\ \bibnamefont {Pufu}}, \ and\
  \bibinfo {author} {\bibfnamefont {S.}~\bibnamefont {Sachdev}},\ }\href
  {\doibase 10.1007/JHEP03(2016)111, 10.1007/JHEP06(2015)037} {\bibfield
  {journal} {\bibinfo  {journal} {JHEP}\ }\textbf {\bibinfo {volume} {06}},\
  \bibinfo {pages} {037} (\bibinfo {year} {2015})},\ \bibinfo {note} {[Erratum:
  JHEP03,111(2016)]},\ \Eprint {http://arxiv.org/abs/1504.00368}
  {arXiv:1504.00368 [hep-th]} \BibitemShut {NoStop}%
\bibitem [{\citenamefont {Chester}\ \emph {et~al.}(2017)\citenamefont
  {Chester}, \citenamefont {Iliesiu}, \citenamefont {Mezei},\ and\
  \citenamefont {Pufu}}]{largeN8}%
  \BibitemOpen
  \bibfield  {author} {\bibinfo {author} {\bibfnamefont {S.~M.}\ \bibnamefont
  {Chester}}, \bibinfo {author} {\bibfnamefont {L.~V.}\ \bibnamefont
  {Iliesiu}}, \bibinfo {author} {\bibfnamefont {M.}~\bibnamefont {Mezei}}, \
  and\ \bibinfo {author} {\bibfnamefont {S.~S.}\ \bibnamefont {Pufu}},\
  }\href@noop {} {\  (\bibinfo {year} {2017})},\ \Eprint
  {http://arxiv.org/abs/1710.00654} {arXiv:1710.00654 [hep-th]} \BibitemShut
  {NoStop}%
\bibitem [{\citenamefont {Jafferis}\ \emph {et~al.}(2017)\citenamefont
  {Jafferis}, \citenamefont {Mukhametzhanov},\ and\ \citenamefont
  {Zhiboedov}}]{zhiboedov}%
  \BibitemOpen
  \bibfield  {author} {\bibinfo {author} {\bibfnamefont {D.}~\bibnamefont
  {Jafferis}}, \bibinfo {author} {\bibfnamefont {B.}~\bibnamefont
  {Mukhametzhanov}}, \ and\ \bibinfo {author} {\bibfnamefont {A.}~\bibnamefont
  {Zhiboedov}},\ }\href@noop {} {\  (\bibinfo {year} {2017})},\ \Eprint
  {http://arxiv.org/abs/1710.11161} {arXiv:1710.11161 [hep-th]} \BibitemShut
  {NoStop}%
\bibitem [{\citenamefont {Chester}\ \emph {et~al.}(2016)\citenamefont
  {Chester}, \citenamefont {Mezei}, \citenamefont {Pufu},\ and\ \citenamefont
  {Yaakov}}]{epsilon}%
  \BibitemOpen
  \bibfield  {author} {\bibinfo {author} {\bibfnamefont {S.~M.}\ \bibnamefont
  {Chester}}, \bibinfo {author} {\bibfnamefont {M.}~\bibnamefont {Mezei}},
  \bibinfo {author} {\bibfnamefont {S.~S.}\ \bibnamefont {Pufu}}, \ and\
  \bibinfo {author} {\bibfnamefont {I.}~\bibnamefont {Yaakov}},\ }\href
  {\doibase 10.1007/JHEP12(2016)015} {\bibfield  {journal} {\bibinfo  {journal}
  {JHEP}\ }\textbf {\bibinfo {volume} {12}},\ \bibinfo {pages} {015} (\bibinfo
  {year} {2016})},\ \Eprint {http://arxiv.org/abs/1511.07108} {arXiv:1511.07108
  [hep-th]} \BibitemShut {NoStop}%
\bibitem [{\citenamefont {Hartnoll}\ \emph
  {et~al.}(2008{\natexlab{a}})\citenamefont {Hartnoll}, \citenamefont
  {Herzog},\ and\ \citenamefont {Horowitz}}]{adscft1}%
  \BibitemOpen
  \bibfield  {author} {\bibinfo {author} {\bibfnamefont {S.~A.}\ \bibnamefont
  {Hartnoll}}, \bibinfo {author} {\bibfnamefont {C.~P.}\ \bibnamefont
  {Herzog}}, \ and\ \bibinfo {author} {\bibfnamefont {G.~T.}\ \bibnamefont
  {Horowitz}},\ }\href {\doibase 10.1103/PhysRevLett.101.031601} {\bibfield
  {journal} {\bibinfo  {journal} {Phys. Rev. Lett.}\ }\textbf {\bibinfo
  {volume} {101}},\ \bibinfo {pages} {031601} (\bibinfo {year}
  {2008}{\natexlab{a}})},\ \Eprint {http://arxiv.org/abs/0803.3295}
  {arXiv:0803.3295 [hep-th]} \BibitemShut {NoStop}%
\bibitem [{\citenamefont {Hartnoll}\ \emph
  {et~al.}(2008{\natexlab{b}})\citenamefont {Hartnoll}, \citenamefont
  {Herzog},\ and\ \citenamefont {Horowitz}}]{adscft2}%
  \BibitemOpen
  \bibfield  {author} {\bibinfo {author} {\bibfnamefont {S.~A.}\ \bibnamefont
  {Hartnoll}}, \bibinfo {author} {\bibfnamefont {C.~P.}\ \bibnamefont
  {Herzog}}, \ and\ \bibinfo {author} {\bibfnamefont {G.~T.}\ \bibnamefont
  {Horowitz}},\ }\href {\doibase 10.1088/1126-6708/2008/12/015} {\bibfield
  {journal} {\bibinfo  {journal} {JHEP}\ }\textbf {\bibinfo {volume} {12}},\
  \bibinfo {pages} {015} (\bibinfo {year} {2008}{\natexlab{b}})},\ \Eprint
  {http://arxiv.org/abs/0810.1563} {arXiv:0810.1563 [hep-th]} \BibitemShut
  {NoStop}%
\bibitem [{\citenamefont {Hartnoll}(2009)}]{adscft3}%
  \BibitemOpen
  \bibfield  {author} {\bibinfo {author} {\bibfnamefont {S.~A.}\ \bibnamefont
  {Hartnoll}},\ }\href {\doibase 10.1088/0264-9381/26/22/224002} {\bibfield
  {journal} {\bibinfo  {journal} {Classical and Quantum Gravity}\ }\textbf
  {\bibinfo {volume} {26}},\ \bibinfo {pages} {224002} (\bibinfo {year}
  {2009})},\ \Eprint {http://arxiv.org/abs/0903.3246} {arXiv:0903.3246
  [hep-th]} \BibitemShut {NoStop}%
\bibitem [{\citenamefont {Herzog}(2009)}]{adscft4}%
  \BibitemOpen
  \bibfield  {author} {\bibinfo {author} {\bibfnamefont {C.~P.}\ \bibnamefont
  {Herzog}},\ }\href {http://stacks.iop.org/1751-8121/42/i=34/a=343001}
  {\bibfield  {journal} {\bibinfo  {journal} {Journal of Physics A:
  Mathematical and Theoretical}\ }\textbf {\bibinfo {volume} {42}},\ \bibinfo
  {pages} {343001} (\bibinfo {year} {2009})},\ \Eprint
  {http://arxiv.org/abs/0904.1975} {arXiv:0904.1975 [hep-th]} \BibitemShut
  {NoStop}%
\bibitem [{\citenamefont {Horowitz}\ and\ \citenamefont
  {Roberts}(2009)}]{adscft5}%
  \BibitemOpen
  \bibfield  {author} {\bibinfo {author} {\bibfnamefont {G.~T.}\ \bibnamefont
  {Horowitz}}\ and\ \bibinfo {author} {\bibfnamefont {M.~M.}\ \bibnamefont
  {Roberts}},\ }\href {\doibase 10.1088/1126-6708/2009/11/015} {\bibfield
  {journal} {\bibinfo  {journal} {JHEP}\ }\textbf {\bibinfo {volume} {11}},\
  \bibinfo {pages} {015} (\bibinfo {year} {2009})},\ \Eprint
  {http://arxiv.org/abs/0908.3677} {arXiv:0908.3677 [hep-th]} \BibitemShut
  {NoStop}%
\bibitem [{\citenamefont {Horowitz}(2011)}]{adscft6}%
  \BibitemOpen
  \bibfield  {author} {\bibinfo {author} {\bibfnamefont {G.~T.}\ \bibnamefont
  {Horowitz}},\ }\enquote {\bibinfo {title} {Introduction to holographic
  superconductors},}\ in\ \href {\doibase 10.1007/978-3-642-04864-7_10} {\emph
  {\bibinfo {booktitle} {From Gravity to Thermal Gauge Theories: The AdS/CFT
  Correspondence: The AdS/CFT Correspondence}}},\ \bibinfo {editor} {edited by\
  \bibinfo {editor} {\bibfnamefont {E.}~\bibnamefont {Papantonopoulos}}}\
  (\bibinfo  {publisher} {Springer Berlin Heidelberg},\ \bibinfo {address}
  {Berlin, Heidelberg},\ \bibinfo {year} {2011})\ pp.\ \bibinfo {pages}
  {313--347},\ \Eprint {http://arxiv.org/abs/1002.1722} {arXiv:1002.1722
  [hep-th]} \BibitemShut {NoStop}%
\bibitem [{\citenamefont {Hartnoll}(2011)}]{adscft7}%
  \BibitemOpen
  \bibfield  {author} {\bibinfo {author} {\bibfnamefont {S.~A.}\ \bibnamefont
  {Hartnoll}},\ }\href@noop {} {\  (\bibinfo {year} {2011})},\ \Eprint
  {http://arxiv.org/abs/1106.4324} {arXiv:1106.4324 [hep-th]} \BibitemShut
  {NoStop}%
\bibitem [{\citenamefont {Dias}\ \emph {et~al.}(2014)\citenamefont {Dias},
  \citenamefont {Horowitz}, \citenamefont {Iqbal},\ and\ \citenamefont
  {Santos}}]{adscft8}%
  \BibitemOpen
  \bibfield  {author} {\bibinfo {author} {\bibfnamefont {{\'O}.~J.~C.}\
  \bibnamefont {Dias}}, \bibinfo {author} {\bibfnamefont {G.~T.}\ \bibnamefont
  {Horowitz}}, \bibinfo {author} {\bibfnamefont {N.}~\bibnamefont {Iqbal}}, \
  and\ \bibinfo {author} {\bibfnamefont {J.~E.}\ \bibnamefont {Santos}},\
  }\href {\doibase 10.1007/JHEP04(2014)096} {\bibfield  {journal} {\bibinfo
  {journal} {Journal of High Energy Physics}\ }\textbf {\bibinfo {volume}
  {2014}},\ \bibinfo {pages} {96} (\bibinfo {year} {2014})},\ \Eprint
  {http://arxiv.org/abs/1311.3673} {arXiv:1311.3673 [hep-th]} \BibitemShut
  {NoStop}%
\end{thebibliography}%
\end{document}